\documentclass{epl}

\shorttitle{Facilitated spin models on Bethe lattice...}
\shortauthor{M. Sellitto, G. Biroli and C. Toninelli}

\title{Facilitated spin models on Bethe lattice: bootstrap
percolation, mode-coupling transition and glassy
dynamics}
\author{Mauro Sellitto\inst{1}, Giulio Biroli\inst{2}, and Cristina
  Toninelli\inst{3}}

\institute{ \inst{1} The Abdus Salam International Centre for
  Theoretical
  Physics, Strada Costiera 11, \\ 34100 Trieste, Italy. \\
  \inst{2} Service de Physique Th{\'e}orique, CEA/Saclay - Orme
  des Merisiers F-91191 Gif-sur-Yvette Cedex, France \\
  \inst{3} Laboratoire de Physique Th{\'e}orique, {\'E}cole Normale
  Sup{\'e}rieure, 24 rue Lhomond Paris, France }

\rec{}{}


\begin{document}

\maketitle

\newcommand{\Tdyn}{T_{\scriptscriptstyle \rm dyn}} 
\newcommand{\Ttap}{T_{\scriptscriptstyle \rm TAP}} 
\newcommand{\Tc}{T_{\scriptstyle \rm c}} 
\newcommand{\tw}{t_{\scriptscriptstyle \rm w}} 
\newcommand{\ec}{e_{\scriptstyle \rm \infty}} 
\newcommand{\eth}{e_{\scriptstyle \rm th}} 
\newcommand{\etap}{e_{\scriptscriptstyle \rm TAP}} 
\newcommand{\teq}{t_{\scriptscriptstyle \rm eq}} 
\newcommand{\tmax}{t_{\scriptscriptstyle \rm max}} 
\newcommand{\chimax}{\chi_{\scriptscriptstyle \rm max}} 
\newcommand{\kB}{k_{\scriptscriptstyle \rm B}} 
\newcommand{\Tg}{T_{\scriptscriptstyle \rm g}} 
\newcommand{\Tmct}{T_{\scriptscriptstyle \rm d}} 
\newcommand{\Eeq}{e_{\scriptstyle \rm eq}} 
\newcommand{\meq}{m_{\scriptscriptstyle \rm eq}} 
\newcommand{\taurel}{\tau_{\scriptstyle \rm rel}} 
\newcommand{\pc}{p_{\scriptscriptstyle \rm c}} 
\newcommand{\phic}{\phi_{\scriptscriptstyle \rm c}}

\begin{abstract} 
  We show that facilitated spin models of cooperative dynamics
  introduced by Fredrickson and Andersen display on Bethe lattices a
  glassy behaviour similar to the one predicted by the mode-coupling
  theory of supercooled liquids and the dynamical theory of mean-field
  disordered systems.  At low temperature such cooperative models show
  a two-step relaxation and their equilibration time diverges at a
  finite temperature according to a power-law. The geometric nature of
  the dynamical arrest corresponds to a bootstrap percolation process
  which leads to a phase space organization similar to the one of
  mean-field disordered systems.  The relaxation dynamics after a
  subcritical quench exhibits aging and converges asymptotically to
  the threshold states that appear at the bootstrap percolation
  transition.
\end{abstract}

\section{Introduction}

Lattice models are widely used in statistical mechanics to gain a
qualitative and often deeper understanding of physical phenomena. In a
seminal work Fredrickson and Andersen (FA)~\cite{FrAn84}
introduced a simple lattice spin model of the liquid-glass transition,
whose Hamiltonian corresponds to uncoupled Ising spins in a positive
magnetic field~\cite{FrBr,Graham}.  The spins represent a
coarse-grained region of the liquid with high ($-1$ spins) or low
($+1$ spins) mobility and the magnetic field, that favors up spins,
leads to very few mobile regions embedded in an immobile background at
low temperature. The spin dynamics is subjected to a {\it kinetic
  constraint}: for each time step a randomly selected spin can flip
only if the number of nearest neighbor down spins is larger or equal
than $f$, where the facilitation parameter $f$ is a number between
zero and the lattice connectivity.  The kinetic constraint mimics at a
coarse grained level the cage effect in super-cooled liquids, where
particles rattle in the cage formed by their neighbors and then move
further if they are able to find a way through the surrounding
particles. At low temperature/high density the latter process is
strongly inhibited leading to an arrest of particle motion over
macroscopic time scales.  Similarly, at high temperature the kinetic
constraint plays a little role and the relaxation is fast, whereas at
low temperature (i.e. high density of up spins) it is hardly satisfied
and relaxation involves strongly cooperative processes which can be
exceedingly slow.  The basic assumption behind this approach is that
the slow dynamics in glass-forming liquids is {\it not} due to a
thermodynamic transition (there is no singularity in the partition
function of non-interacting Ising spins in a magnetic field), but
rather to a purely dynamic mechanism.

Kinetically constrained models (KCM), such as the FA and the KA
model~\cite{KoAn}, encode in a very compact way many features of real
glass-forming systems.  Stretched exponential relaxation~\cite{FrBr},
super-Arrhenius equilibration time~\cite{Graham}, dynamical
heterogeneity~\cite{Ha} and self-diffusion/viscosity
decoupling~\cite{BeChGa} are some examples.  They also display basic
glassy features, such as physical aging~\cite{KuPeSe} and effective
temperature~\cite{Se}, first derived in the out of equilibrium
dynamics of random p-spins systems~\cite{BoCuKuMe}.  In finite
dimensions no transition takes place at finite temperature (or
non-unit density)~\cite{Reiter,ToBiFi1}, while on Bethe lattices a
dynamical transition similar to that predicted by the mode coupling
theory (MCT) may occur~\cite{ReMaJa, PiYoAn, ToBiFi1}.

The nature of this similarity is far from being obvious. On one hand,
KCM are defined by {\it ad hoc} kinetic rules which, although
physically motivated, have little to do with the microscopic
(Newtonian or Brownian) dynamics of glass-forming liquids. On the
other hand, most results concerning the aging dynamics were derived
within fully connected spin-glasses, which unlike KCM have built-in
quenched disorder and a non-trivial thermodynamics at low temperature
(characterised by a replica-symmetry broken phase).  In order to make
further progress in this direction we investigate the relaxation
dynamics of the FA model on a Bethe lattice.  (In a recent work
motivated by similar issues~\cite{Sz}, MCT has been applied to 1D FA
model). We show in detail that the mechanism of ergodicity breaking
can be understood in terms of {\it bootstrap percolation} (as already
mentioned in Refs.~\cite{Reiter, ReMaJa, PiYoAn, ToBiFi1}), and that
the phase space organization and several dynamical features are well
described by the predictions of MCT and its out of equilibrium
generalisation~\cite{BoCuKuMe}.

\section{FA model on Bethe lattice and the bootstrap percolation} 
\label{bethe} 
The FA model we consider in this paper consists of $N$ Ising spin
variables, $\sigma_i=\pm1$, $i=1, \dots, N$, on a Bethe lattice with
connectivity $k$.  The denomination Bethe lattice refers here to a
random regular $k$-graph, that is a graph taken at random within the
set of graphs with fixed connectivity $z=k+1$ on each site. In this
random graph all sites are on the same footing, and there is no
``surface'' effect. The Hamiltonian is simply: $ H =
-\frac{1}{2}\sum_i\sigma_i$, and the spin dynamics is subjected to a
kinetic constraint: at each time step a randomly chosen spin is
flipped with the rate: $ w (\sigma_i\rightarrow
-\sigma_i)={\mbox{min}} \left\{ 1,\, {\rm e}^{-\beta \sigma_i}
\right\} $, {\it only if } the number of nearest neighbor in state
$-1$ is larger than or equal to $f$ ($\beta=1/T $ is the inverse
temperature).  FA~models with $f>1$ are usually called cooperative
since their relaxation time grows faster than the Arrhenius law at low
temperature, at variance with the simpler non-cooperative case $f=1$.
In the literature it seems to be known, even if not discussed in
detail, that the FA model on a Bethe lattice has an
ergodic/non-ergodic transition at a finite temperature for
$k>f>1$~\cite{ReMaJa, PiYoAn}.  This is closely related to the
bootstrap percolation transition, as we will explain in the following.

The local structure of a Bethe lattice is Cayley-tree like with $k$
branches going up from each node and one going down.  Let us call $B$
the probability that, without taking advantage of the configuration on
the bottom, the spin $\sigma_i$ is in the state $-1$ or can be brought
in this state rearranging the sites above it. $B$ verifies a simple
iterative equation:
\begin{eqnarray} 
B &=& (1-p) + p \sum_{i=0}^{k-f} {k \choose i} B^{k-i} (1-B)^{i} 
\label{B} 
\end{eqnarray} 
where $ p= 1/(1+ {\rm e}^{-1/T} )$ is the probability that the spin is
$+1$ in thermal equilibrium.  Eq.~(\ref{B}) has, for small $p$ (i.e.
high-temperature) only the solution $B=1$: with probability one each
spin can be flipped in the up-state using a certain (finite) number of
allowed moves. At low temperature, when $p$ is large enough, a
transition occurs to a fixed point $B<1$ for any choice of $k>f>1$
(the critical value of $p$ depending on $k$ and $f$).  The properties
of this blocking (or jamming) transition can be easily established.
For arbitrary small $1-B$ the eq.~(\ref{B}) becomes $(1-B) = p {k
  \choose f-1} (1-B)^{k-f+1}+ \dots$ Since the power of $1-B$ on the
right and left hand side are different only a discontinuous transition
is possible for $k>f$. One can also show that if $1-p$ (i.e. the
temperature) is small enough then a transition has to take place.
Furthermore at the transition it is easy to check that $1-B$ has a
square root singularity coming from low temperatures. This mixed
character of first and second order is similar to the behavior of the
non-ergodicity parameter in MCT.  Notice also that there is no
transition for the non-cooperative case $f=1$ (no matter the value of
$k$).  Let us now explain how the above transition is related to the
bootstrap percolation (BP),~\cite{BP,ChLeRe}.  In BP 
each site of a lattice is first occupied by a particle at random with
probability $p$.  Then, one randomly remove particles which have less
than $m=k-f+2$ neighbors.  Iterating this procedure leads to two
possible asymptotic results~\cite{ChLeRe}. If the initial particle
density is larger than a given threshold $\pc$
there is a residual infinite cluster of particles that remains at the
end of this procedure, whereas for $p<\pc$
the density of residual particles is zero.  In ref.~\cite{ChLeRe} it
has been shown that the probability $P$ that a particle is blocked
because it has more than $k-f+1$ neighboring particles above it which
are blocked (without taking advantage of vacancies below) satisfies
the eq. $P = p \sum_{i=0}^{f-1} {k \choose i} P^{k-i} (1-P)^{i} $.
Changing variable $i'=k-i$ and the notation particle-vacancy/up-down
spin one obtains that $1-P$ verifies the same equation of $B$.  Thus
the two transitions coincide.  In fact, if $P>0$ then there will be
particles (spins) blocked permanently and $B<1$. If $P=0$ then there
exists, with probability one, a sequence of allowed moves that brings
a random equilibrium configuration to the configuration with all sites
$-1$ by definition (this set of configurations has been called the
high temperature partition by Fredrickson and Andersen).  Hence, each
site can flip to $-1$ after a certain number of allowed moves and
$B=1$.  Although irreducibility is not equivalent to ergodicity we
expect that, because of the trivial thermodynamic measure, the
bootstrap and dynamical transition coincide.
Of special interest is the fraction $\phi$ of spins that are permanently
frozen. It can be computed exactly from $P$~as:
\begin{eqnarray} \label{pb} 
  \phi &=&  p \sum_{i=0}^{f-1} {k+1 \choose i} P^{k+1-i}
  (1-P)^{i} + (1-p) \sum_{i=0}^{f-1} {k+1 \choose i} (p h_{k,f})^{k+1-i}
  ( 1-p h_{k,f} )^i  \,,
\end{eqnarray} 
where $h_{k,f}=\sum_{i=0}^{f-2} {k \choose i} P^{k-i}(1-P)^{i}$.  The
two contributions are respectively the probability that a spin is
frozen in the state $\pm 1$. In the following we analyze the
case $f=2$, $k=3$, which mimics the square lattice with $f=2$.  In
this case $P = p \left(P^{3}+3P^{2} (1-P) \right)$, and 
%
%
the critical value of $p$ at which the transition takes place is
$\pc=\frac{8}{9}$ which gives $\Tc=1/\ln \frac{\pc}{1-\pc} =
0.480898$.  The corresponding fraction of frozen spins is $\phic =
\phi(\Tc) \simeq 0.67309$.

\begin{figure}[h] 
  \twofigures[width=7cm]{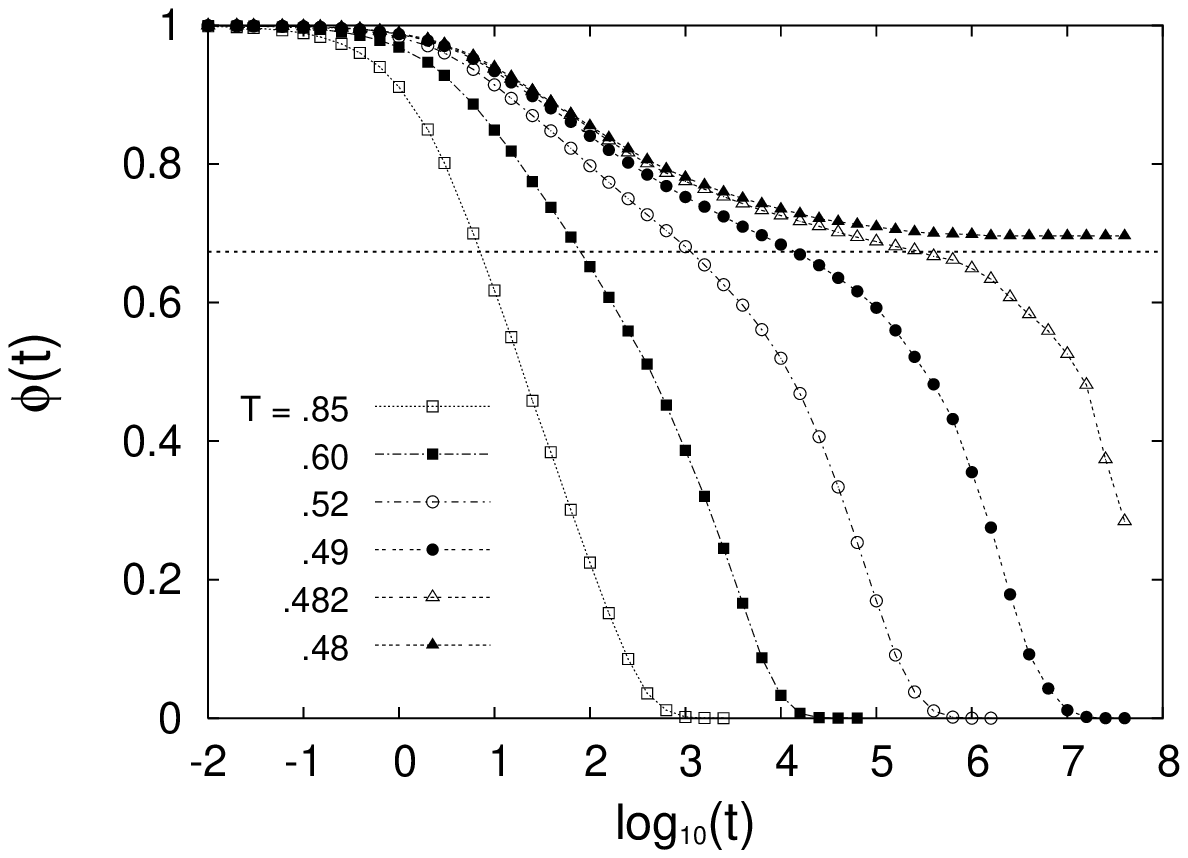}{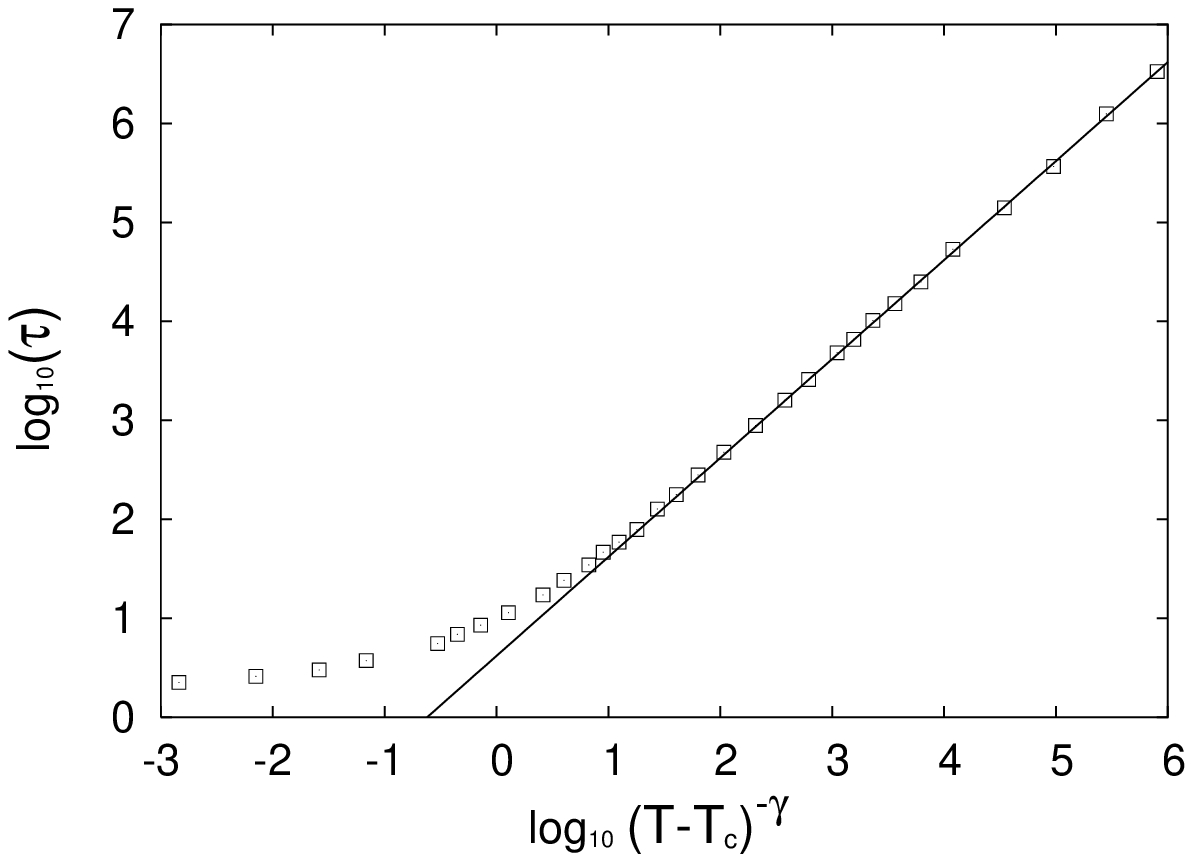}
\caption{Persistence function $\phi(t)$ vs time $t$ during the
  equilibrium dynamics at temperatures $T$, for a FA model with
  facilitation parameter $f=2$, on a Bethe lattice with connectivity
  $k+1=4$.  The straight line is the plateau $\phic$ predicted by the
  bootstrap percolation argument. System size $N=2^{18}$.}
  \label{fig_phi_grafo} 
\caption{ Integral relaxation time $\tau$  vs temperature $T$.  The
  dynamical critical temperature $\Tc$ is the one predicted by the
  bootstrap percolation argument, $\Tc \simeq 0.481$. The power-law
  exponent is $\gamma \simeq 2.9$.}
  \label{fig_tau}
\end{figure}

\section{Equilibrium dynamics and relaxation time} 
\begin{figure}[h] 
  \twofigures[width=7cm]{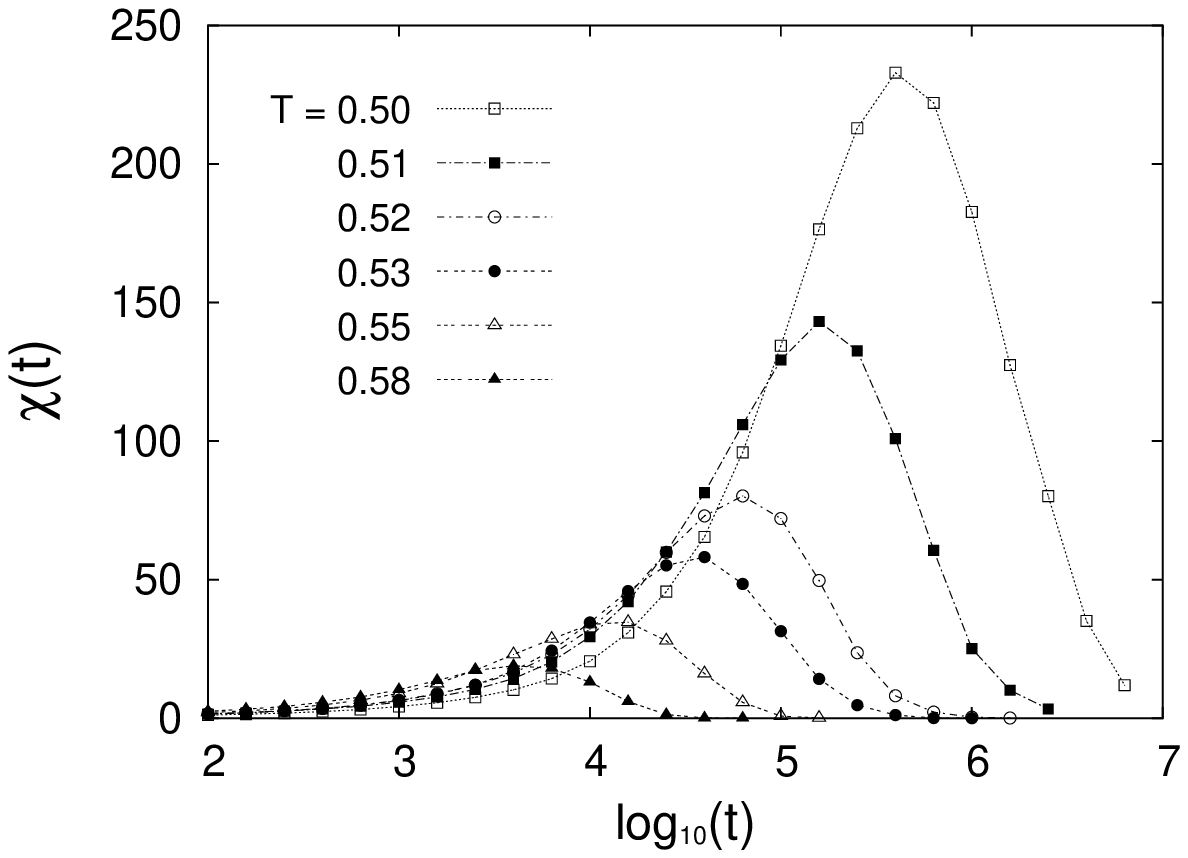}{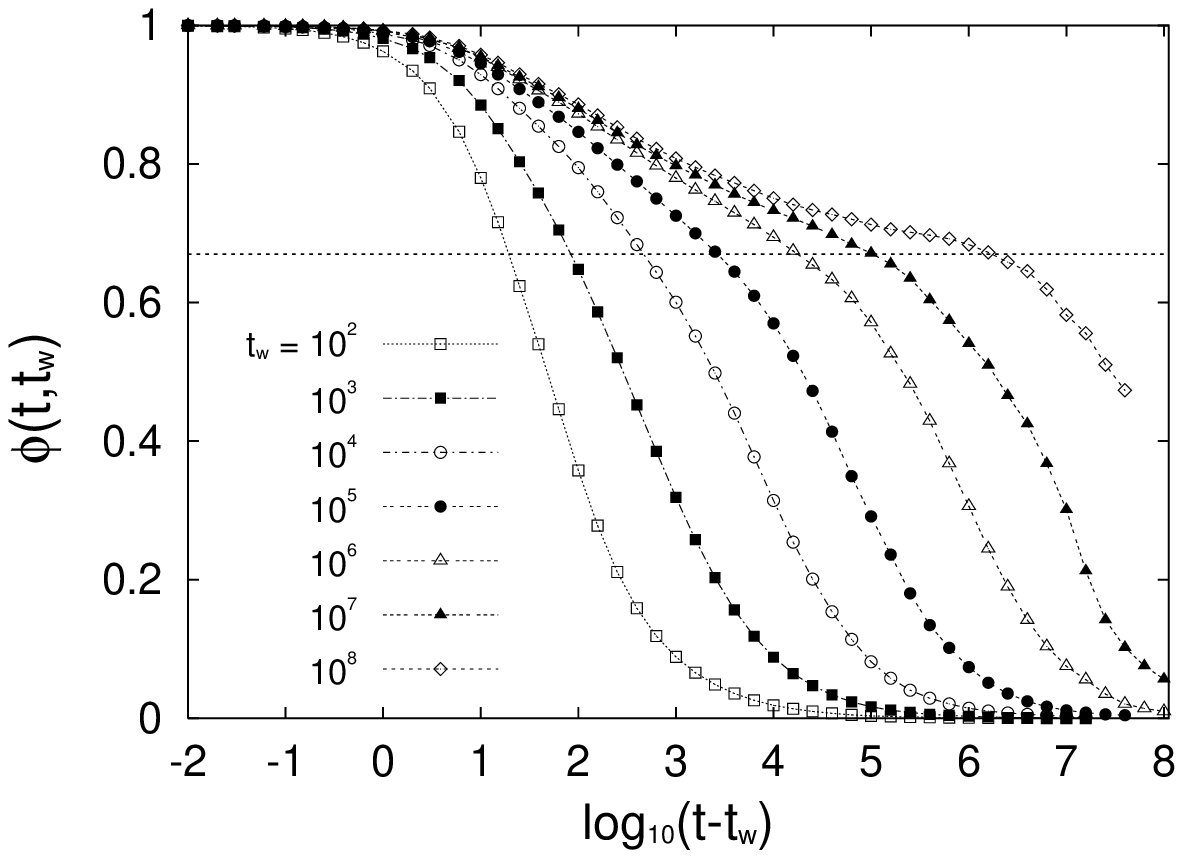}
\caption{Equilibrium dynamical susceptibility $\chi(t)$ vs time $t$
  at temperature~$T$. System size $N=2^{14}$.}
  \label{fig_chi} 
\caption{Two-time persistence
  $\phi(t,\tw)$ after a subcritical quench at temperature $T=0.4$.
  The straigth line is the (temperature-independent) plateau $\phic$
  predicted by the bootstrap percolation argument.}
  \label{fig_aging}
\end{figure}

The equilibrium dynamics of KCM is usually analyzed in terms of the
persistence function $\phi(t)$, i.e. the probability that a spin has
never flipped between times $0$ and $t$~\cite{RiSo}. The analytical
results obtained above for the fraction of frozen spins are directly
relevant to the long-time limit of $\phi(t)$ that indicates when the
ergodicity is broken:
\begin{eqnarray} 
  \lim_{t \to \infty} \phi(t)  =  0 \,, \,{\rm for} \,\,\,\, T > \Tc \,; \qquad
  \lim_{t \to \infty} \phi(t)  > 0 \,, \,{\rm for} \,\,\,\, T \le \Tc  
\label{phi_oo} 
\end{eqnarray} 
and therefore it plays the same role of the non-ergodicity parameter
in supercooled liquids, and the Edwards-Anderson parameter in
spin-glasses.  In order to compare the above results with numerical
simulation we have implemented a faster than the clock Monte Carlo
code~\cite{Werner} for the dynamics of the FA model on a Bethe
lattice. In fig.~\ref{fig_phi_grafo} we show the persistence function
$\phi (t)$ for $f=2$, and $k=3$.  The data are obtained by using an
initial configuration with equilibrium magnetization $m =
\tanh(\beta/2)$ which is by construction a typical equilibrium
configuration.  Upon decreasing $T$ towards $\Tc$, $\phi(t)$ develops
a clear two-step behavior in which the density of frozen spins slowly
approaches a plateau, and then decay to zero on longer time scale.
Similar results are obtained for the spin-spin correlation function.
The plateau value is equal to the fraction of spins that never flip,
see eq.~(\ref{pb}).  The estimated values of $\Tc$ and $\phic$ are in
good agreement with the theory.  We have also obtained the relaxation
time, $\tau(T)$, as the integral of the persistence function.
Consistently with the prediction $\tau(T)$ diverges at a finite
temperature exactly given by $\Tc \simeq 0.48$.  Moreover, we find
that the divergence has a power-law singularity $\tau(T) \sim
(T-\Tc)^{-\gamma}$, with an exponent $\gamma \simeq 2.9$, see
fig.~\ref{fig_tau}. This behavior is observed over five decades of
relaxation time and holds for any temperature below~1. We also mention
that at low temperature, the departure from the plateau is well
described by the von Schweidler law, while the late stage of
relaxation (the so-called $\alpha$ regime) obeys the time-temperature
superposition principle, $\phi(t) = \Phi(t/\tau(T))$, the scaling
function $\Phi$ being fitted well by a Kohlrausch-Williams-Watts
stretched exponential.  Since the BP has features of both first and
second order transitions we expect increasing fluctuations on
approaching $\Tc$. Such fluctuations have a purely dynamical nature as
being related to the dynamical order parameter $\phi(t)$ and can be
detected by measuring the dynamical susceptibility $\chi(t) =
N(\langle \phi^2\rangle - \langle \phi\rangle^2)/T $. We observe
indeed that $\chi(t)$ develops a peak (see fig.~\ref{fig_chi}) that
diverges as $\tau(T)$ for $T\rightarrow \Tc$, similarly to what has
been found in random p-spin model and MCT~\cite{FrPa,BiBo}.

%
%

\section{Energy relaxation and the threshold states} 

To better understand the glassy behaviour of the FA model on Bethe
lattice we have studied the aging dynamics and the phase-space
organization below $\Tc$.  For comparison, it is useful to recall the
scenario of mean-field disordered models of structural
glasses~\cite{BoCuKuMe,BaBuMe}.  Below a temperature $\Tmct$
ergodicity is broken, and the configuration space decomposes in an
exponential number (in the system size) of ergodic components or TAP
states (corresponding to different minima of the local magnetization
free energy functional, i.e. the TAP free energy).  The logarithm of
the number of states, the configurational entropy, jumps from zero to
a finite value at $\Tmct$. The TAP states which dominate the Boltzmann
weight at $\Tmct$ are called {\it threshold states} and play an
important role in the dynamical behaviour. Relaxational dynamics at
$T<\Tmct$, starting from a random configuration exhibits aging: the
system approaches asymptotically the threshold states being unable to
thermalise within any TAP state at temperature $T$.  One time
observables, like the energy, converge asymptotically to the
corresponding threshold value, while two-time correlation and response
functions do not obey the usual equilibrium fluctuation-dissipation
relation.  This scenario has relevance beyond the physics of glassy
systems: a striking example is the typical-case analysis of hard
combinatorial optimization problems~\cite{hard}.  Note however that
disordered systems may have an aging dynamics dominated by different
layers of TAP states depending on the cooling schedule~\cite{MoRi}.
In the following we check if the scenario outlined above is relevant
to the FA model.  The counterpart of a TAP state in the FA model will
be the set of all configurations having the same backbone of frozen
spins.

\begin{figure}[h] 
  \twofigures[width=7cm]{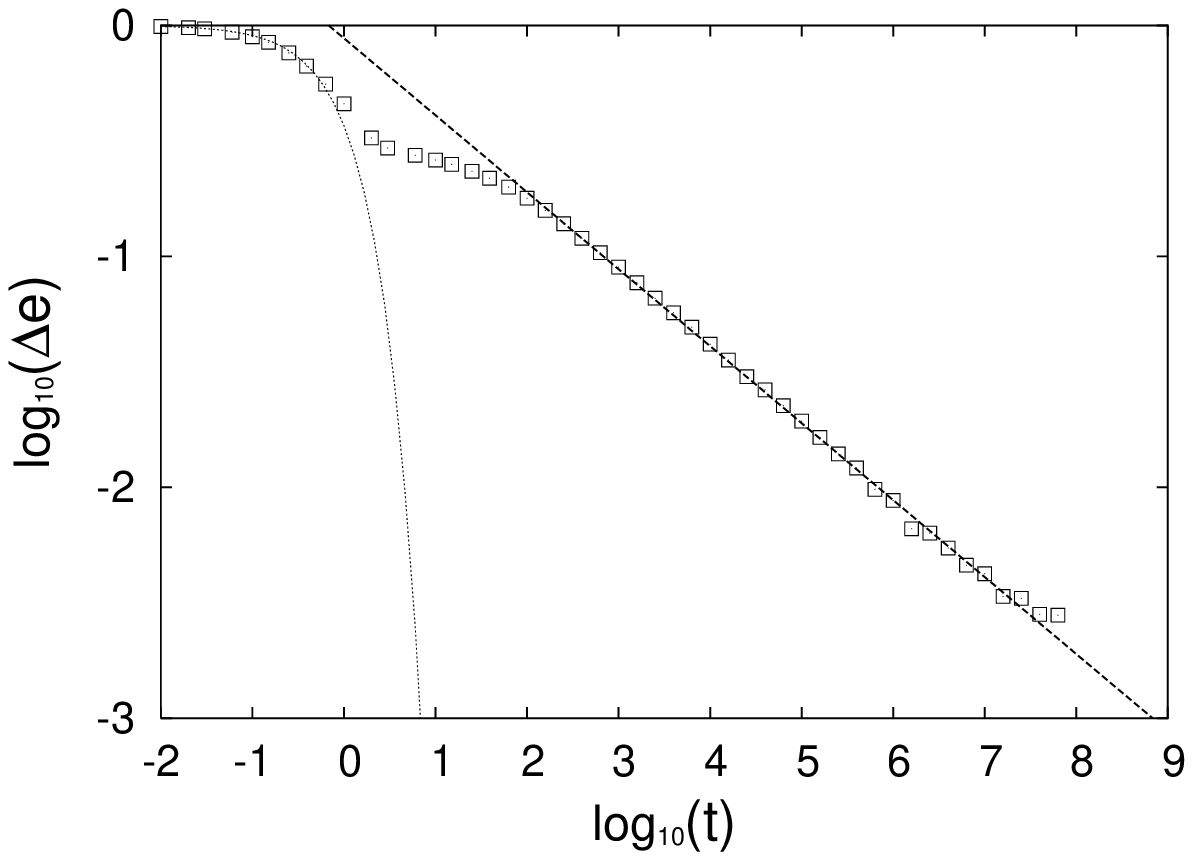}{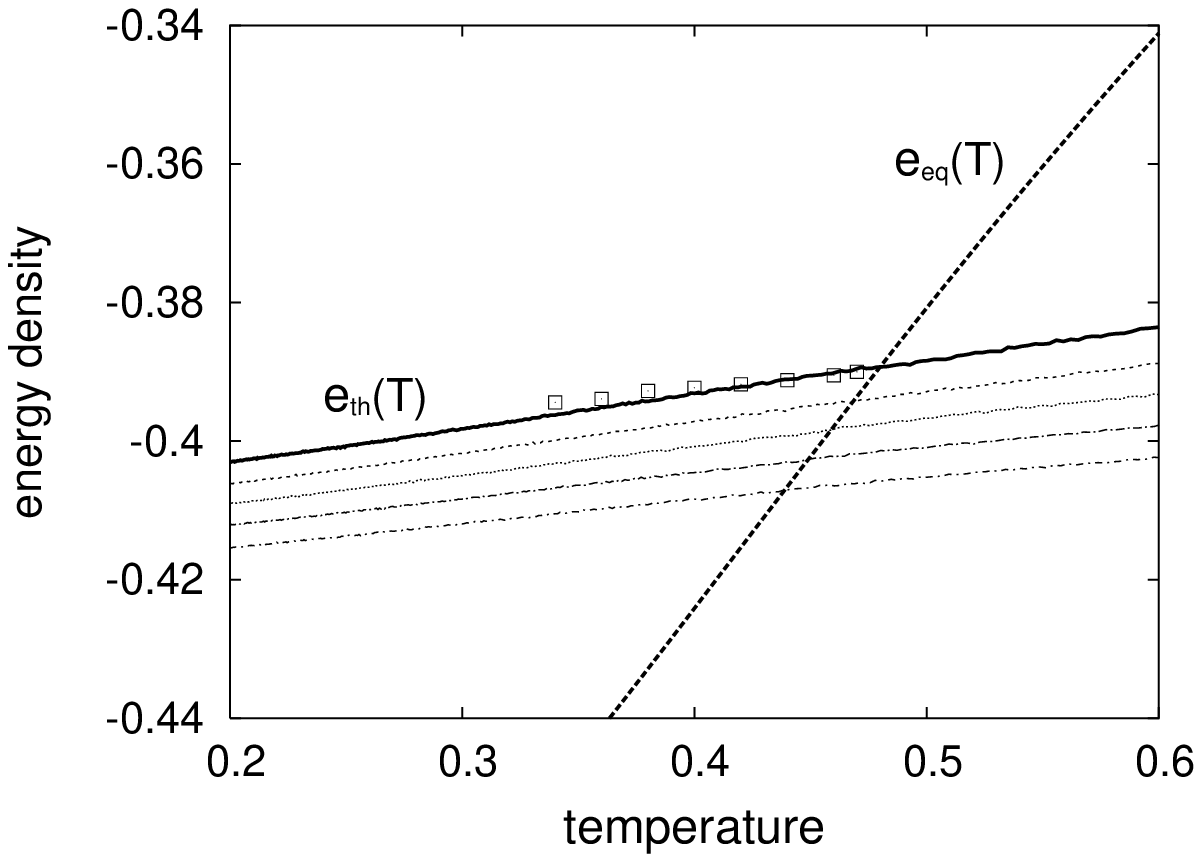}
\caption{Excess energy density $\Delta e(t) = (\ec(T) - e(t))/(\ec(T)
  - e(0))$ vs time $t$, after a subcritical quench at temperature
  $T=0.38$. The dashed line is a power-law $\Delta e(t) \sim t^{-\nu}$
  with $ \nu \simeq 0.33$. The dotted line is a
  temperature-independent exponential relaxation.}
 \label{fig_M_c4f2}
\caption{The dashed line is the equilibrium energy $\Eeq(T) =
  -\tanh(1/2T)/2$. The full line is the threshold energy $\eth(T)$
  obtained by cooling (heating) a configuration equilibrated at $\Tc
  \simeq 0.48$. Square symbol represents the asymptotic energy,
  $\ec(T)$, of a subcritical quench at temperature~$T$. Four
  subthreshold states are also shown; they are obtained by cooling
  (heating) an equilibrium configuration at temperature
  $T=0.47,\,0.46,\,0.45,$ and $0.44$.}
\label{fig_e_c4f2}
\end{figure}

We generally find that after a subcritical quench at temperature $T$
the energy approaches an asymptotic value, $\ec(T)$, above the
equilibrium one, $\Eeq(T)=-\tanh(1/2T)/2$. Accordingly, the two-time
persistence between time $\tw$ and $t$ shows a two-step aging
behavior, see fig.~\ref{fig_aging}.  In fig.~\ref{fig_M_c4f2}
we plot the excess energy $\Delta e(t) = (\ec(T) - e(t))/(\ec - e(0))$
for $T=0.47$.  After a short-time temperature-independent exponential
relaxation, $\Delta e(t)$ decays according to a power-law $\Delta e(t)
\sim t^{-\nu} $.  A similar behavior is observed for different
quenching temperatures, with $\nu$ varying in the range $[0.31,
\,0.34]$ for $T \in [0.34,\,0.48]$.  
%
%
%
In fig.~\ref{fig_e_c4f2} we compare $\ec(T)$, with the energy density
of the threshold states $\eth(T)$. The latter is obtained by cooling
(heating) at a slow enough rate $\dot{\epsilon}$ a random equilibrium
configuration at $\Tc$.  In this way the system reaches equilibrium
within one of the ergodic components and its energy at temperature $T$
represents, in the limit $\dot{\epsilon} \rightarrow 0$, the threshold
energy.  We find that for $T$ not too far from $\Tc$ the results
obtained with the two protocols are equal within the numerical
accuracy, see fig.~\ref{fig_e_c4f2}. For $T \ll \Tc$, we cannot
exclude a small discrepancy between $\eth(T)$ and $\ec(T)$ due to the
difficulty to access the asymptotic dynamics in the available time
window.  This equality is by no means trivial and strongly suggests
that the threshold states dominate indeed the off-equilibrium
behavior. Subthreshold states with $\etap(T) < \eth(T)$ are sampled by
considering an initial equilibrium configuration at $T<\Tc$.  These
states represent different ergodic components and are completely
separated from each other in the dynamical evolution. They can be
followed by varying the temperature and their phase space organization
is very similar to the one of p-spins models (for comparison see
fig.~1 in ref.~\cite{BaBuMe}).  The main difference is that TAP states
in the FA model are stable against thermal fluctuations: upon heating
they never disappear (i.e.  $\Ttap=\infty$ in the notation of
ref.~\cite{BaBuMe}) This is a consequence of the athermal nature of
the blocking transition, which in our case is due to a purely kinetic,
rather than geometric, constraint.  We find blocked structures even
above $\Tc$ but they have a vanishing Boltzmann weight and are not
seen by the off-equilibrium dynamics.

\section{Conclusion}\label{conclusion}

We have studied facilitated spin models of glassy dynamics on a Bethe
lattice with fixed connectivity $k+1$ and facilitation parameter $f$.
For the non-cooperative case, $f=1$, there is no transition at
non-zero temperature, the relaxation time is Arrhenius, and no
two-step relaxation is observed. The cooperative case, $k>f>1$, is
much richer: at a finite temperature $\Tc$ a giant cluster of frozen
spins appears and hence ergodicity is broken.  The transition presents
a mixed character: the fraction of frozen spins jumps discontinuously
at $\Tc$ with a square root singularity, and therefore there are
critical fluctuations with diverging correlation length and time
scales.  The relaxation time diverges as a power-law at $\Tc$, and a
two-step relaxation is observed both in and out of equilibrium.  The
geometric mechanism behind such a jamming transition is a bootstrap
percolation phenomenon, as pointed out
previously~\cite{ReMaJa,PiYoAn,ToBiFi1}. The onset of jamming in
sphere packings and its relation with the bootstrap or $k$-core
percolation~\cite{k-core}, and rigidity percolation~\cite{RP} has been
recently discussed in ref.~\cite{Liu}.  Microscopically this
transition is rather different from the dynamic transition occurring
in mean field disordered systems~\cite{Rivoire}.  In spite of this, we
have shown that several characteristics of the glassy dynamics studied
above are remarkably similar to those found in the mode-coupling
theory and the aging dynamics of mean-field models of structural
glasses.  This seems to be related to the emergence of a very similar
organization of the phase space, in which the sets of configurations
having the same backbone of frozen spins play the role of TAP states.
In finite dimension the Bethe lattice glass transition is replaced by
a possibly sharp crossover, as recently shown for the KA
model~\cite{ToBiFi1}.  Finite-size systems also exhibit a blocking
transition at a finite temperature, due to the exceedingly slow
approach to the thermodynamic limit in bootstrap percolation
problems~\cite{BP,Ken}.  Numerical evidences suggest that some
peculiar features of mean-field structural glasses remain even in
finite dimension, at least on long but finite
time-scales~\cite{Se,BaKuLoSe}.  Further work would be certainly
valuable to improve the understanding of this crossover.

\section{Acknowledgement} 
We thank S. Franz for interesting discussions and his participation to
the early stage of this work. GB and CT thank D.S. Fisher for
interesting discussions. 
MS thanks M. Weigt for discussions about $k$-core percolation on random graphs.
CT is supported by the EU contract HPRN-CT-2002-00319 (STIPCO).

\stars

\end{document}